\documentclass[pra,aps,showpacs,epsf,twocolumn]{revtex4}

\usepackage{graphicx}
\usepackage{epsfig}
\usepackage{amsmath}
\usepackage{amssymb}
\usepackage{multirow}
\usepackage{color}

\newcommand{\me}{\rm e}
\newcommand{\mi}{\rm i}
\newcommand{\abs}[1]{\left\vert#1\right\vert}
\newcommand{\bra}[1]{\langle #1|}
\newcommand{\ket}[1]{|#1\rangle}

\begin{document}
\title{Interacting atoms in optical lattices}
\author{Johan Mentink and Servaas Kokkelmans}
\affiliation{Eindhoven University of Technology, P.O.~Box~513, 5600~MB  Eindhoven, The Netherlands}
\date{\today}

\begin{abstract}

We propose an easy to use model to solve for interacting atoms in an optical lattice. This model allows for the whole range of weakly to strongly interacting atoms, and it includes the coupling between relative and center-of-mass motion via anharmonic lattice terms. We apply this model to a high-precision spin dynamics experiment, and we discuss the corrections due to atomic interactions and the anharmonic coupling. Under suitable experimental conditions, energy can be transferred between the relative and center-of-mass motion, and this allows for creation of Feshbach molecules in excited lattice bands.
\end{abstract}


\maketitle

Optical lattices form a suitable environment for high precision experiments with interacting atoms. Two atoms can be isolated from other atoms by placing them on a single lattice site, and many sites can be filled simultaneously. While the lattice parameters such as depth and geometry can be adjusted via the laser field, the interactions can be tuned using a Feshbach resonance by applying an external magnetic field. Feshbach molecules in a lattice, created by sweeping the magnetic field over resonance, can be transferred into deeper vibrational bound states, for instance, by applying stimulated Raman adiabatic passage~\cite{Winkler}.

Precise values for relative interaction strengths of rubidium atoms were derived by studying coherent collisional spin dynamics in an optical lattice~\cite{Widera2005,Widera2006}. These high precision measurements provide challenges for theoretical coupled-channels models based on current state-of-the-art interaction potentials~\cite{vanKempen2002}. One may wonder for instance, at what level of precision it is possible to calculate interaction properties before the Born-Oppenheimer approximation, which is the underlying foundation for the potentials, breaks down.

Before conclusions can be drawn on the limitations of two-body interaction models, one has to make sure that the correct comparison is made between theoretical quantities of such a calculation, and the measurements that depend on these interactions. For instance, one cannot always put the resulting scattering lengths of a two-body collision, defined in the limit of zero collision energy, as the on-site interaction in a Hubbard model. The divergence of the scattering length on resonance will give rise to physically unrealistic large energy shifts. A resonant interaction takes the two-body interaction in the unitarity limit, where the solution of scattering wavefunctions are shifted over $\pi/2$ compared to non-interacting atoms, and one would rather use expressions based on the (energy) dependent scattering phase shift. This argument also applies for high-precision experiments on non-resonant systems, since small energy-dependent corrections already can be of importance. Also, the relative and center-of-mass motion of two interacting atoms, which can become coupled due to different atomic species~\cite{Bertelsen,Deuretzbacher} and anharmonic terms in the lattice potential~\cite{Deuretzbacher}, can give rise to shifts in the on-site interaction.

The model we put forward in this paper is conceptionally simple and easy to use. It is based on first-order perturbation theory starting from an existing solution of two interacting atoms in a harmonic potential. We show that the model is valid for moderately deep lattices, i.e. where tunneling to neighbouring sites is negligble. In this way, we are able to make a proper comparison between the high-precision measurements by Widera et al.~\cite{Widera2005} and accurate rubidium scattering models. We demonstrate the importance of energy-dependence in the scattering phase shift and of anharmonic corrections for this experiment, and also show how experiments using a Feshbach resonance can produce molecules of a mixed relative and center-of-mass motion nature.

This paper is outlined as follows: First we give a description of our model. Then we apply it to a spin-dynamics experiment in a lattice. In the third section, we discuss the nature of the molecules that have a mixed relative and center-of-mass motion, predicted by our model. We end with a concluding section.

\section{Model}
In the ultracold regime only s-wave interactions are allowed, and we conveniently model the interaction with a pseudo potential~\cite{huang}
\begin{equation}
V_{\rm int}(r)=\frac{4\pi\hbar^2}{m}\,a\, \delta^{(3)}(r)\frac{\partial}{\partial r}r.
\end{equation}
Here $r$ is the distance between two colliding atoms, and $a$ the s-wave scattering length~\cite{footnote}. The use of this zero-range pseudo potential is allowed, since the range of the real interaction potentials is much smaller than the lattice spacing. The total Hamiltonian for a pair of interacting atoms in an optical lattice is the described by
\begin{eqnarray}
\lefteqn{H = -\frac{\hbar^2}{2m}\left(\frac{1}{2}\nabla_\mathbf{R}^2 + 2\nabla_{\bf r}^2 \right) + }\\ \nonumber
& & \hspace{1cm} + V_{\rm lat}\left({\bf R}+\frac{\bf r}{2}\right) + V_{\rm lat}\left({\bf R}-\frac{\bf r}{2}\right) + V_{\rm int}(r).
\end{eqnarray}
Here
\begin{equation}
V_{\rm lat}({\bf x})=V_0\sum_{j=1}^{3}\sin^2(k_{\rm L}x_j)
\end{equation}
is the optical lattice potential with depth $V_0$ and spacing $d=\pi/k_{\rm L}$, where is $k_{\rm L}$ the wavenumber corresponding to the laser frequency and ${\bf x}=(x_1,x_2,x_3)$ is decomposed along the lattice axes. ${\bf R}$ and ${\bf r}$ denote the center-of-mass (CM) and relative position of the atoms. The restriction to a pair of interacting atoms is an approximation, but exact for deep lattices (for example in the Mott-insulater regime) for two trapped atoms per lattice site. Separation of the CM and relative motion is possible when each site is treated as an harmonic oscillator (HO) with frequency $\omega=\sqrt{2V_0k_{\rm L}^2/m}$. The lattice potential for two particles can then be written as

\begin{equation}
 V_{\rm lat}\left({\bf R}+\frac{\bf r}{2}\right) + V_{\rm lat}\left({\bf R}-\frac{\bf r}{2}\right) \approx \frac{1}{2}m\omega^2(2{\bf R}^2+\frac{1}{2}{\bf r}^2).
\end{equation}
An exact solution for two interacting atoms in a HO trap can be derived for the relative motion, as was first shown by Busch {\it et.~al} \cite{Busch1998}. However, lattice effects beyond the harmonic approximation, i.e.~anharmonic terms as well as tunneling to neighbouring sites, can be of significant importance even when the atoms are considered trapped in a single site~\cite{Widera2006,Deuretzbacher}. The consequence of the introduction of such terms is that the CM and relative motion become coupled. To tackle this problem, we handle anharmonic and tunneling terms in a perturbative procedure, starting from the separable solution:
\begin{equation}
\psi^{(0)}_{Ss}({\bf R},{\bf r})=\Phi_S({\bf R})\phi_s({\bf r}).
\end{equation}
Here $\Phi_S({\bf R})$ and $\phi_s({\bf r})$ are the exact solutions of the CM and relative motion, corresponding to states labeled with $S=\{S_1,S_2,S_3\}$ and $s=\{n,l,m\}$, respectively. We use different quantum numbers for CM and relative motion, to adapt optimally to both the symmetry of the perturbation term and the symmetry of the interaction region. Therefore, CM is always described in cartesian coordinates, with quantum numbers $S_j=0,1,2,..$ for the different lattice axes. Relative motion is decomposed in spherical coordinates, with principal quantum number $n=1,2,3,..$, and orbital quantum numbers $l$ and $m$. Without the presence of atomic interactions, i.e. when~$a=0$, many HO states are degenerate. The corresponding collection of quantum numbers is denoted with the symbol $\{\cdot,\cdot,\cdot\}$.

\begin{figure}[t]
\includegraphics[width=\columnwidth]{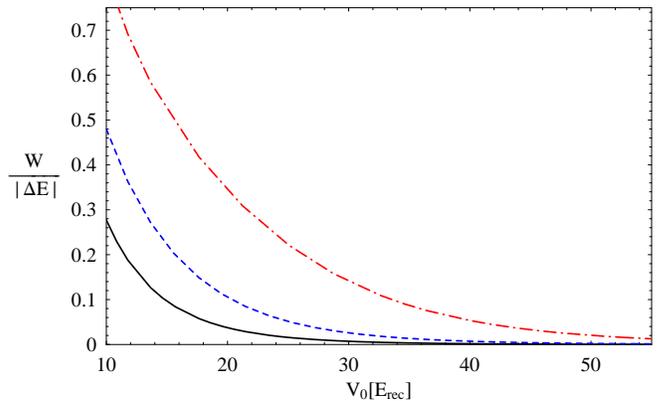}
\caption{\small Estimation of tunneling and anharmonic effects based on exact lattice solutions for non-interacting particles. The ratio of band width $W$ and level shift $\Delta E$ is shown as function of lattice depth $V_0$, given in units of recoil energy $E_{\rm rec}=\hbar^2k_{\rm L}^2/2m$. The black solid and red dash-dotted lines correspond to the lowest and second but lowest symmetric bands, respectively. The blue dashed line represents the lowest anti-symmetric band. \label{f:tunnelanharm3}}
\end{figure}

For the case~$a=0$, one can also find an exact lattice solution completely in cartesian coordinates, for example in terms of Mathieu functions on single-particle coordinates. This allows us to estimate the importance of tunneling compared to anharmonic terms, shown in Fig.~\ref{f:tunnelanharm3}. Tunneling gives rise to a broadening $W=E_t-E_b$ of the HO level structure, which results in the band structure of a square lattice~\cite{Ashcroft}, and also has a level shift $\Delta E=E_{\rm HO}-(E_t+E_b)/2$. Here $E_t$ and $E_b$ are the band top and band bottom energies, respectively, and $E_{\rm HO}$ is the corresponding HO level.

\begin{figure}[t]
\center{\includegraphics[width=\columnwidth]{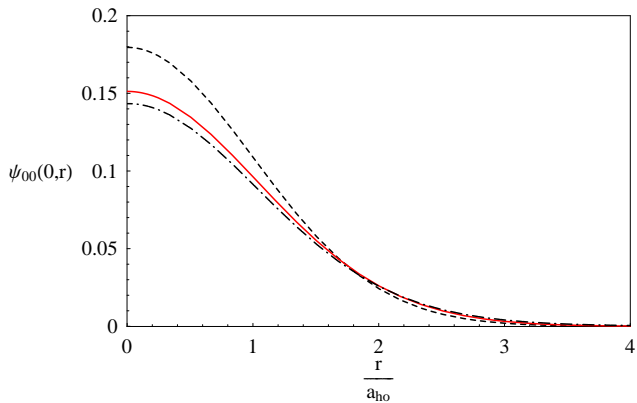}}
\caption{\small Comparison of the exact and perturbative wavefunctions in the lowest band, for $a=0$. Shown are the zeroth order HO solution (dashed line), the first order corrected solution (red solid line) and the exact localized Wannier function (dash-dotted line), for CM coordinate $R=0$. The relative coordinate is scaled on $a_{\rm ho}$, with $a_{\rm ho}=\sqrt{2\hbar/(m\omega)}$ the harmonic oscillator length of relative motion.}
\label{f:checkwavefunc}
\end{figure}

We find that sufficiently deep in the Mott phase, in particular for lattice depths as used in current experiments, that anharmonic terms dominate above tunneling: $W \ll |\Delta E|$. Since also $|\Delta E| \ll E_{\rm HO}$, we expect in the case of non-zero interatomic interactions, a perturbative approach with anharmonic corrections only is sufficient to yield accurate results. In the remainder, we therefore restrict ourselves to single-site solutions including anharmonic terms. As perturbation term we then have:
\begin{eqnarray}
H^\prime({\bf R},{\bf r}) & = & V_{\rm lat}\left({\bf R}+\frac{\bf r}{2}\right) + V_{\rm lat}\left({\bf R}-\frac{\bf r}{2}\right) + \\ \nonumber
& & \hspace{2.1cm} -\frac{1}{2}m\omega^2(2{\bf R}^2+\frac{1}{2}{\bf r}^2).\label{e:perturbterm}
\end{eqnarray}
We define our perturbative solutions in accordance with standard perturbation theory:
\begin{equation}
E_{Ss}=E^{(0)}_{Ss} + \bra{\psi^{(0)}_{Ss}}H^\prime\ket{\psi^{(0)}_{Ss}} + \ldots,
\end{equation}
and
\begin{equation}\label{e:firstorderwavefunc}
\psi_{Ss}=\psi^{(0)}_{Ss} + \sum_{\genfrac{}{}{0pt}{}{S^\prime\ne S}{s^\prime\ne s}}\frac{\bra{\psi^{(0)}_{S^\prime s^\prime}}H^\prime\ket{\psi^{(0)}_{Ss}}}{E^{(0)}_{Ss}-E^{(0)}_{S^\prime s^\prime}}\psi^{(0)}_{S^\prime s^\prime} + \ldots,
\end{equation}
for the first order corrected energies and wavefunctions, respectively. $E^{(0)}_{Ss}=E_S+e_s$ is the sum of the HO energies corresponding to the CM and relative motion. To check the accuracy of our model, we compare the results with the exact lattice solution available for the case without interaction. Regarding the energy, we compare $E_{Ss}$ with $E_{\rm exact}=(E_t+E_b)/2$. With only first order corrections we already have $(E_{Ss}-E_{\rm exact})/E_{\rm exact}<0.7\%$, for the lowest band at a lattice depth $V_0=25E_{\rm rec}$. Note that the accuracy is of the same order as the relative contribution of tunneling (Fig.~\ref{f:tunnelanharm3}), and therefore the meaning of second order terms is limited. However, this first-order accuracy is sufficient for our description of the spindynamics experiment in Section III. Naturally, the accuracy improves when lattice depth is increased. We also compared the results for the wavefunction, and an example is given in Fig.~\ref{f:checkwavefunc}, for the same value of the lattice depth. Shown are the first order corrected wavefunction, together with the zeroth order HO solution and the Wannier function, being the exact solution for a single site. Anharmonic corrections make the trap less tight compared to the HO approximation, resulting in a small decrease of the probability for finding the particles near $r=0$ around the origin, corresponding to a small increase of probability density in the barrier. The difference between the Wannier function and the HO wavefunction with anharmonic corrections is small compared to the difference between the latter and the unperturbed HO wavefunction. This shows that the method converges quickly.

When we include interactions ($a\ne 0$), the zeroth order relative wavefunction and energy depend parametrically on the scattering length according to the Busch model \cite{Busch1998}: $\phi_s({\bf r})=\phi_s({\bf r};a)$ and $e_s=e_s(a)$. Hence the anharmonic correction becomes a function of scattering length as well. Note however that the perturbation term itself does not depend on the scattering length, illustrating that the interaction is modeled exactly.

Finally, we show how to construct a solution for the whole lattice. Due to the relative interactions, there is no periodicity for the relative motion. However, the CM motion remains periodic. In fact the CM boundary conditions are dictated by the Bloch-theorem. These boundary conditions are satisfied when we construct the lattice solution as
\begin{equation}
\psi_{\textbf{Q}Ss}({\bf R},{\bf r})=\sum_{\bf D}\,\me^{\mi {\bf Q}\cdot{\bf D}}\psi_{Ss}({\bf R}-{\bf D},{\bf r}),
\end{equation}
with direct lattice vector ${\bf D}=(n_1d,n_2d,n_3d)$, integers $n_i$, and ${\bf Q}$ the CM quasi momentum. This demonstrates that the wavefunctions $\psi_{Ss}$ obtained with our model are effectively Wannier functions of a pair of interacting atoms in an optical lattice.

\section{Spindynamics}
In this section we apply our model to the spindynamics experiment, as carried out in the Bloch group \cite{Widera2005,Widera2006}. This allows us to investigate systematically the effects of anharmonic terms and a non-zero interatomic interaction.

In an optical lattice one is able to trap several spin-states at the same time. Starting with atoms prepared in a specific one-particle (hyperfine) spin-state $\ket{f,m_f}$, collisions between two such atoms gives access to other two-particle spin configurations. When only weak magnetic fields are applied, the total magnetization is conserved, and therefore coherent collisions between two-particle states of equal total two-particle spin $F$ occur. This can be described by a Rabi-like model. For atoms prepared in $f=1$, which is the case we will treat here, this is only a two level system, with effective Rabi frequency

\begin{equation}
\Omega'_{\rm if}=\sqrt{\Omega^2_{\rm if}+\delta^2_{\rm if}}.
\end{equation}
Here $\Omega_{\rm if}$ is the bare Rabi frequency depending on the coupling strength of the spin-changing collision. Detuning $\delta_{\rm if}$ contains two contributions:

\begin{equation}
\delta_{\rm if}=\delta_0+\delta(B^2).
\end{equation}
$\delta_0$ is given by the difference of two interaction energies corresponding to collisions that leave the spin configuration unchanged, and $\delta(B^2)$ is a second order Zeeman shift between the initial and final states. By performing the experiment at different magnetic field strengths $B$, the Bloch group was able to substract $\Omega_{\rm if}$ and $\delta_0$, thereby being able to derive precise values for relative interaction strengths of rubidium. Treating the interaction energy as a linear perturbation in the parameter $k_{\rm L}a$, one can derive the Rabi parameters as
\begin{equation}
\Omega_{\rm if}=\frac{2\sqrt{2}}{3\hbar}\tilde{U}(a_2-a_0), \quad \delta_0=\frac{1}{3\hbar}\tilde{U}(a_2-a_0).
\end{equation}
The factor $\tilde{U}$ is here defined as
\begin{equation}
\tilde{U}=\frac{4\pi\hbar^2}{m}\times\int{\rm d}^3x\abs{\psi}^4,
\end{equation}
with $\psi$ the lowest HO eigenfunction. $\tilde U$ depends on the lattice depth, but is independent of the scattering length. Hence in this approach, differences between interaction energies are due to scattering-length differences only. The values of the scattering lengths $a_F$, $F=0,2$, corresponding to collisions in subchannel $F$, are calculated based on highly accurate rubidium potentials~\cite{vanKempen2002}, and are given in Table~\ref{t:scatteringlengths}. Note that $\psi$ can deviate significantly from the above-mentioned solution of two interacting atoms in a trap, and therefore it can be expected that a proper treatment of interatomic interactions could have a large impact on the Rabi frequency. While this effect will be mostly pronounced when close to a Feshbach resonance, an effect can also be expected when highly accurate measurements are performed, as in this experiment of Widera {\it et al}.

The experimental results for the scattering lengths, based on the above Rabi model~\cite{Widera2005}, agreed just within error bars with the predictions. This apparent discrepancy was most clearly seen for the $f=1$ case. We will now investigate this $f=1$ experiment by calculating the interaction strengths from our lattice model, and analyze the effects of anharmonic terms and exact interatomic interactions. Note that anharmonic corrections were also taken into account in Ref.~\cite{Widera2006}, which already led to a better agreement between theory and experiment. We can most clearly compare theory and experiment by regarding the frequency $\Omega'^{\rm HO}_{\rm if}$ at $B=0$, since the field dependence does not depend on the interatomic interactions.

\begin{table}[b]
\caption{\small Theoretical predictions for the scattering lengths of the $F=0$ and $F=2$ channel, for atoms with one-particle spin $f=1$, based on accurate rubidium interaction potentials~\cite{vanKempen2002}. Here $F$ is the total two-particle spin. The values are given in units of the Bohr radius ${\rm a}_0$.}
\center{
 \begin{tabular}{|c|c|}\hline
 $F$ & $a_F$ [${\rm a}_0$]\\ \hline
 $0$ & $101.78\pm0.2$\\
 $2$ & $100.40\pm0.1$\\ \hline
 \end{tabular}}
\label{t:scatteringlengths}
\end{table}

As a starting point for comparison, we apply the Rabi model as described above, with $\tilde{U}$ calculated from HO solutions, and we use the scattering lengths from Table~\ref{t:scatteringlengths} to obtain the effective Rabi frequency. This gives $\Omega'^{\rm HO}_{\rm if}(B=0) = 2\pi \times 49.27\,\rm{Hz}$. Note that for a comparison between theory and experiment, it is sufficient to consider $\Omega'_{\rm if}$ at zero magnetic field, since the magnetic field dependence does not involve any knowledge of the interatomic interactions. Then, we first calculate $\Omega'_{\rm if}(B=0)$ by using Wannier functions in the expression for $\tilde{U}$, in order to analyze the effects of anharmonic terms only. Wannier functions are exact solutions for a lattice without interactions. This is similar to the procedure performed by Widera~{\it et al.} \cite{Widera2006}. Second, we want to analyse the effect of having exact interatomic interactions only, and calculate the Rabi parameters by using the solutions for two interacting atoms in a harmonic trap:

\begin{eqnarray}
\Omega_{\rm if}=\frac{2\sqrt{2}}{3\hbar}\left(E_{\rm int}(a_2)-E_{\rm int}(a_0)\right) \\
\delta_0=\frac{1}{3\hbar}\left(E_{\rm int}(a_2)-E_{\rm int}(a_0)\right),
\end{eqnarray}
where
\begin{equation}
E_{\rm int}(a)=E(a)-E(a=0).
\end{equation}
Here we define the total energy $E(a)=E^{(0)}_{Ss}(a)=E_S+e_s(a)$, with $S=\{0,0,0\}$ and $s=\{3,0,0\}$, according to the solution of two interacting atoms in a trap~\cite{note1}. Third, the same is done, but also with the anharmonic terms included by taking $E(a)=E_{Ss}(a)$, in order to compute the combined effect. The results are shown in Table~\ref{t:corrections}, by calculating the ratio of the different $\Omega'_{\rm if}(B=0)$ over the initial frequency $\Omega'^{\rm HO}_{\rm if}(B=0)$. The table also shows a comparison with the experimentally obtained effective Rabi frequency. In all calculations a lattice depth of $V_0=45E_{\rm rec}$ is used. Note that the third calculation is the most precise one, with exact interactions, giving rise to modified wavefunctions and energy levels compared to the HO calculation. Moreover, the anharmonic effects are also included via perturbation theory up to a high precision, which follows from the table (discussion could also go to the end).
\begin{table}[b]
\caption{\small (left column) Correction factor $\Omega'_{\rm if}/\Omega'^{\rm HO}_{\rm if}$ of the effective Rabi frequency $\Omega'_{\rm if}$, compared with the same quantity computed in the HO approximation with the interaction treated as linear perturbation. The first row shows the result when only anharmonic effects are taken into account. The result in the second row is obtained with only higher order interaction effects taken into account. In the third row the results are given of both effects acting together. Right column: corresponding values for the ratio between the theoretical $\Omega'_{\rm if}$ and the experimentally obtained effective Rabi frequency $\Omega'^{\rm exp}_{\rm if}=2\pi\times(35.4\pm0.7){\rm Hz}$.}
\center{
 \begin{tabular}{|r|ll|}\hline
 included corrections & $\Omega'_{\rm if}/\Omega'^{\rm HO}_{\rm if}$ & $\Omega'_{\rm if}/\Omega'^{\rm exp}_{\rm if}$\\ \hline
 anharmonic & 0.897 & 1.25 \\
 interatomic interactions & 1.033 & 1.43 \\
 anharmonic + interatomic interactions & 0.935 & 1.30 \\ \hline
 \end{tabular}}
\label{t:corrections}
\end{table}

From the results we find that anharmonic terms induce a negative shift of order 10 \%. To the contrary, higher order interaction effects induce a positive shift, which is of order 3\%. Hence, anharmonic corrections are dominant, while the first order approximation for the interaction energy is already fairly accurate. The net result is a 7\% improvement with respect to the initial model of the Bloch group. Although the theoretical and experimental values still differ by 30\%, this is just within theoretical and experimental error bars. Note that the largest contribution to the theoretical error bar is due to the small difference $a_2-a_0$, which is only a percent of the values of $a_0,a_2$ themselves. A second issue is a possible systematic error in $\tilde{U}$, related to the uncertainty in the lattice depth. At present there is no direct measurement of this coupling constant.

\section{Adiabatic creation of Feshbach molecules with a CM motion}

In the previous section we applied our model to non-resonant scattering of atoms in the lowest lattice band. This means that the energies of the relative and CM motion are relatively close to their respective HO ground states. The coupling between CM and relative motion is taken into account, which also includes coupling to higher bands when second order lattice perturbation terms are incorporated. However, the interaction induced coupling between the bands is negligible since the energy distance to non-ground state CM motion is too large. When the interaction strength is increased, this is no longer true, and several relative and CM states become (nearly) degenerate, and therefore first order lattice perturbation terms will already result in an efficient coupling of these states. In this section, we show that this implies that atoms can be transferred into molecules, with a simultaneous transfer of quantized energy of the relative motion to the CM motion, and vise versa.

\begin{figure}[t]
\center{\includegraphics[width=\columnwidth]{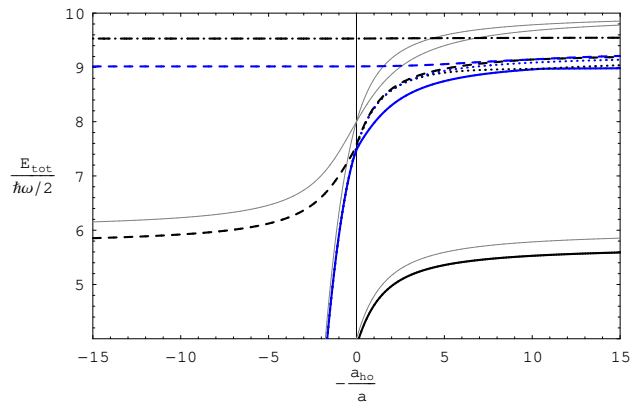}}
\caption{\small (Color online) Total energy of the combined relative and CM system, as a function of the reciprocal scattering length. The upper thick lines represent the six states that are coupled by anharmonic terms. Thin gray lines denote the zeroth order HO levels. The scattering length is scaled on $a_{\rm ho}$, with $a_{\rm ho}=\sqrt{2\hbar/(m\omega)}$ the harmonic oscillator length of the relative motion.}
\label{f:coupledlevels}
\end{figure}

Tuning the scattering length through resonance can give rise to a transfer of atoms from one HO level to the next one~\cite{Busch1998}, and under adiabatic conditions this can be observed experimentally \cite{Stoferle2006}. In a pure harmonic trap, one is therefore able to increase the relative energy by $2\hbar\omega$. This is illustrated in Fig.~\ref{f:coupledlevels} by the thin gray lines. The system can become degenerate to other states which, in turn, can be lifted by anharmonic terms. In general, coupling strengths depend on both scattering length and lattice depth. Within first order perturbation theory, states of equal total quantum number $S_1+S_2+S_3+n$ are coupled. To find the first-order energy corrections, we have to diagonalize the matrix whose elements are given by
\begin{equation}
\bra{\psi_{Ss}^{(0)}}H_{HO}+H^\prime\ket{\psi_{S^\prime s^\prime}^{(0)}}=\delta_{SS^\prime}\delta_{ss^\prime}E^{(0)}_{Ss} + \bra{\psi^{(0)}_{Ss}}H^\prime\ket{\psi^{(0)}_{S^\prime s^\prime}}
\end{equation}
Here $\delta_{ii^\prime}$ is the Kronecker delta. Without interaction ($a\to0^-$) these states are degenerate as well, while for $a$ being small and positive the interaction induced coupling becomes negligible, as was the case in the previous section.

We exploit the symmetry of the perturbation term in order to analyze which relative and CM states will be coupled. This can be done conveniently for the case $a=0$. Since the spatial symmetry of the wavefunction is not changed by the interaction, the same states will be coupled for $a\ne0$, however, with coupling constants depending on $a$. The perturbation term is symmetric in cartesian coordinates, hence coupling occurs only between states that have equal symmetry along the lattice axes. For $a\to0^-$ and without perturbation terms, the first excited energy consists of several degenerate states $|\psi_{Ss}\rangle$ which can be divided into two sets. The first set consists of a CM ground state and excited relative states with labels $S=\{0,0,0\}$ and $s=\{3,l,m\}$, and the second set consist of excited CM states and a relative ground state with $S=\{2,0,0\}$ (and cyclic permutations), and $s=\{0,0,0\}$. In the first case, quantum numbers can be $l,m=\{0,0\},\{2,0\},\{2,-2\},\{2,+2\}$, hence four relative states in total. As required, the number of relative states that are coupled reduces to three when we take proper linear combinations of d-waves. So dictated by the symmetry of the anharmonic terms, six states are coupled in total. Note that it is quite remarkable that a coupling to relative d-wave motion is possible, without an interatomic coupling on short range. This coupling between s-waves and d-waves is indirect. The s-waves are coupled to excited CM states which in turn are coupled to d-waves, illustrating the long-range character of the anharmonic coupling.

\begin{figure}[t]
\center{\includegraphics[width=\columnwidth]{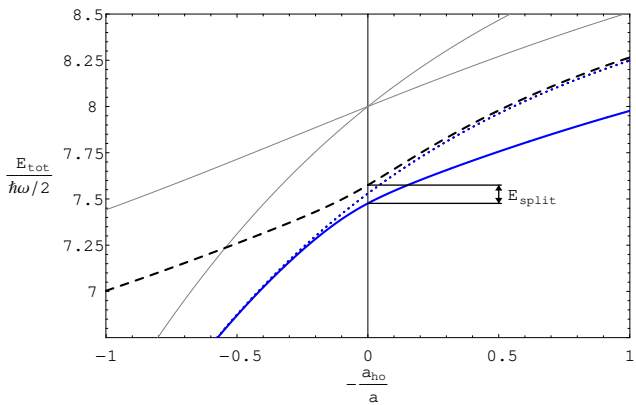}}
\caption{\small (Color online) Zoom in of Fig.~\ref{f:coupledlevels} around resonance ($1/a=0$). The scattering state (black dashed line) is coupled to one excited CM state (blue solid line), which contains a molecular bound state for positive $a$, while the other two states (dotted line) are degenerate and remain uncoupled around resonance. The splitting energy is indicated by $E_{\rm split}$.}
\label{f:crossing}
\end{figure}

The main result of this section is shown in Fig.~\ref{f:coupledlevels}. Total energy for the combined relative and CM motion is shown as a function of the scattering length. Thick lines represent the six states that are coupled by anharmonic terms. Thin gray lines indicate the corresponding zeroth order HO levels. Also the perturbed uncoupled ground state is shown, using a thick solid line. The calculation is done for a lattice depth of $V_0=25E_{\rm rec}$. Fig.~\ref{f:coupledlevels} shows the presence of d-wave states in the upper two solutions, indicated with black dash-dotted and blue dashed lines. These solutions have only weak dependence on the s-wave scattering length, owing to the indirect coupling via the excited CM states. Below the d-wave states a dashed black solution is shown, which connects asymptotically to two consecutive levels, similar to the results of the Busch model for a s-wave scattering state. However, the three remaining solutions (dotted black, dotted blue and solid blue) are of a different nature. These reflect the presence of excited CM states, and asymptotically connect to the molecular state of the relative motion for positive $a$. A zoom-in around resonance ($1/a=0$) is given in Fig.~\ref{f:crossing}. It can be clearly seen that the coupling between relative and CM motion, caused by the long-range anharmonic terms, gives rise to a transfer of energy between these two motions. On resonance only the dashed and solid lines are coupled, giving rise to an energy splitting indicated by $E_{\rm split}$. The two dotted solutions are coupled at small negative values of the scattering length and with much smaller splitting energies.

It is interesting to consider possible applications of this anharmonic coupling at long range, and interatomic coupling at short range. This can be done by exploiting different timescales when changing the scattering length. The scattering length can be changed by utilizing the magnetic field dependence of the scattering length via a Feshbach resonance. One can transfer for instance atoms from the lowest band into the next band by ramping the magnetic field, and by slowly ramping back associate molecules with an excited CM motion. This would result in molecular energy levels that deviate significantly from the energy of ground state molecules labeled with $S=\{0,0,0\}$ and $s=\{1,0,0\}$. Such higher molecular levels correspond to molecules in different (partly-filled) Brillouin zones, which should be possible to detect~\cite{Koehl}. We note that these excited CM molecules, compared to ground state molecules, have a larger spatial extent in the CM motion.

The typical timescale $\tau$ for slowly ramping back is given by $\tau \gg \hbar/E_{\rm split}$. In contrast to other lattice induced molecules, see {\it e.g.} \cite{Winkler2006}, these excited CM molecules do not arise from tunneling to neighboring sites, but from the anharmonic shape of a single lattice site. In addition, these excited CM molecules could be observable even for rather deep lattices ($V_0>50E_{\rm R}$), since the anharmonic effects decay only weakly with increasing lattice depth, whereas tunneling effects decay exponentially.

\section{Conclusion}

In conclusion, we proposed an easy to use method to solve for interacting atoms in optical lattices, where the relative and center-of-mass motion of the two interacting atoms are coupled via the anharmonic terms of the lattice. The interactions are treated exactly using a boundary condition rising from a pseudo potential. The anharmonic terms of the lattice potential are treated as a perturbation of the exact solution for two cold interacting atoms in a harmonic trap. We applied this model to the Mainz spin dynamics experiment \cite{Widera2005,Widera2006} for $f=1$. The interaction energy is computed as the difference between two-atom energy levels with and without interaction. This model gives a more rigorous interpretation of the experiment compared to previous descriptions, in terms of two-body scattering properties. We find that the derived scattering lengths agree within the experimental and theoretical error bars. Apart from applying our model to spin dynamics, we are in a good position to analyze future optical lattice experiments where the interactions are made very strong by utilizing Feshbach resonances. Strong interactions can induce coupling between the relative and center-of-mass motion, which allows for an energy exchange between these two motions, and which can be used to produce (Feshbach) molecules with an excited center-of-mass motion. This model can also be used as a starting point for a description of photoassociation in an optical lattice near a Feshbach resonance~\cite{Jaksch,Junker}.

This work was supported by the Netherlands Organization for Scientific Research (NWO).


\begin{thebibliography}{99}
\bibitem{footnote} The scattering length can be replaced by a function of the relative momentum, in order to get an advanced energy-dependence of the scattering phase shift. In case of a constant $a$, one accounts for a trivial energy dependence $\delta(k)=\arctan ka$, which is valid for broad Feshbach resonances and non-resonant scattering.
\bibitem{note1} The lowest state at positive scattering length is the molecular bound state. Because we aim to describe scattering between atoms here, we use the one-but-lowest symmetric state, with $s=\{3,0,0\}$.
\end{thebibliography}
\end{document}